\documentstyle[preprint,aps,prd]{revtex}
\draft
\begin{document}
\title
{Thomas-Fermi Description of Incoherent Light Scattering  
from an Atomic-Trap BEC}

\author
{Eddy Timmermans and Paolo Tommasini}
\address{Institute for Theoretical Atomic and Molecular Physics}
\address{Harvard-Smithsonian Center for Astrophysics}
\address{Cambridge, MA 02138}

\date{\today}
\maketitle
\begin{abstract}

        We present a Thomas-Fermi treatment of resonant incoherent
scattering of low-intensity light by a dilute spatially confined
Bose-Einstein condensate.   The description gives simple analytical results
and allows scattering data
from finite-size condensates to be interpreted in terms of the properties of
the homogeneous BEC-system.
As an example, we show how the energy dispersion of the elementary excitations
can be measured from
scattering by a finite-size atomic-trap condensate.
As a second example, we point out that
a near-resonant scattering experiment can observe quasi-particle creation
caused by particle annihilation.

\end{abstract}


\narrowtext

For most applications of the atomic-trap Bose-Einstein condensates 
\cite{Ket} -- \cite{Hul}, it will be beneficial to have a highly populated 
condensate, and recent experiments have succeeded in increasing the
number of condensed atoms \cite{Ketlat}. 
It is fortunate then, that in this limit
(for bosons interacting
through an inter-particle potential of positive scattering length) 
many quantities can be calculated analytically.
The simplifying assumption is the Thomas-Fermi approximation 
\cite{Leg}--\cite{hung},
which presumes that the local behavior of the BEC is similar to that
of a homogeneous BEC with a chemical potential equal to the local effective 
chemical potential 
$\mu ({\bf r}) = \; \mu_{T} - V({\bf r}) $, where $\mu_{T}$ is the chemical 
potential of the trap system and 
$V({\bf r})$ the trapping potential.  

	In this letter, we describe 
incoherent light scattering in a Thomas-Fermi approximation.
The result, with a proper understanding of its limits,
offers insight that will be useful
in interpreting experimental scattering spectra. The 
formalism describes the scattering event as scattering of light by
quasi-particles and includes the effects of recoil and intermediate state 
energy neglected in the off-resonant limit reported by Javaneinen 
\cite{Jav1},\cite{Jav2}.

	The dynamical Thomas-Fermi approximation of this paper
consists of describing the fluctuations that cause an incoherent
scattering event to occur near a position ${\bf r}$, by the corresponding
fluctuations in a homogeneous system of chemical potential $\mu ({\bf r})$.
In this picture, 
the differential cross section for incoherent scattering, 
$d^{2} \sigma_{\rm{inc}} / d\Omega d\omega$, where $d \Omega$ represents
the infinitesimal solid angle and $d \omega$ the infinitesimal energy range
($\hbar = 1$, in our units) 
over which the scattered particles are detected, reduces to 
an integral over a cross section density :
\begin{equation}
\frac{d^{2} \sigma_{\rm{inc}}}{d\Omega d\omega} \approx
\int d^{3} r \left[ 
\frac{d^{2} \sigma_{\rm{inc,hom}} / d\Omega d\omega}{V} \right]
_{\mu = \mu ({\bf r})} \; \; , 
\label{e:tfinc}
\end{equation}
where $d^{2} \sigma_{\rm{inc,hom}} / d\Omega d\omega$ is the cross section
for incoherent scattering from a macroscopic, homogeneous system of volume V and
chemical potential $\mu ({\bf r})$. 

	We use a second quantized
representation in which $c_{\bf k}, c^{\dagger}_{\bf k}$ denote the 
annihilation
and creation of a single-atom state 
with the atom in the atomic ground state (the trapping 
state) and its center-of-mass in the plane wave state of wave vector
${\bf k}$; $\tilde{c}_{\bf k}, \tilde{c}^{\dagger}_{\bf k}$ denote
the corresponding annihilation and creation operators for atoms in the excited
(resonant) atomic state.
In the resonant scattering process, the interaction with the electric 
field of the incident light, $E_{in} \; \hat{\epsilon}_{in} \; \exp
(i \left[ {\bf k}_{in} \cdot {\bf r} - \omega_{in} t \right])$,
where
$E_{in}$ represents the intensity and 
$\hat{\epsilon}_{in}$
the polarization of the incident light,
promotes an atom to its excited atomic state.
The excited atom subsequently deexcites, creating a photon of
electric field $E_{0} \; \hat{\epsilon}_{out} \; \exp
(-i \left[ {\bf k}_{out} \cdot {\bf r} - \omega_{out} t \right])$, where
$E_{0}$ is the single-photon
intensity
(in Gaussian units, $E_{0} = \sqrt{ 2 \pi kc/ V}$, 
where $k$ is the resonant photon wave number).
The corresponding absorption and emission Hamiltonian operators in the
interaction picture are :
\begin{eqnarray}
\hat{H}^{\rm{(abs)}} (t) &=& - E_{in} \; ({\bf d} \cdot \hat{\epsilon}_{in} )
\; \;
\exp (-i \omega_{in} t) \sum_{\bf k} \tilde{c}^{\dagger}_{{\bf k} + 
{\bf k}_{in}} (t) c_{\bf k} (t) \; \; , \nonumber
\\
\hat{H}^{\rm{(em)}} (t) &=& - E_{0} \; 
({\bf d} \cdot \hat{\epsilon}_{out} )^{\ast} \;
\exp (i \omega_{out} t) \sum_{{\bf k}'} c^{\dagger}_{{\bf k}' - 
{\bf k}_{out}} (t) \tilde{c}_{{\bf k}'} (t) ,
\label{e:inth}
\end{eqnarray}
where ${\bf d}$ denotes the resonant atomic dipole moment.  

	The amplitude
of a state $|F \rangle $ of the scattering system, 
a time T after the system 
was in its initial state $|in \rangle$, is, to lowest order in
perturbation theory, equal to :
\begin{equation}
{\cal A}(|in\rangle , |F\rangle ;T) = 
- \int_{-T/2}^{T/2} dt \int_{0}^{T/2+t} d\tau
\langle F| \hat{H}^{\rm{(em)}} (t) \hat{H}^{\rm{(abs)}} (t-\tau) |in \rangle 
\; ,
\label{e:amp}
\end{equation}
where the photon absorption occurs a time $\tau$ earlier than the photon
emission \cite{rwa}.  
In the limit of low-intensity incident light, 
each photon is scattered by a system
of ground state atoms.
Indicating the many-body vacuum of excited--state atoms, $|\tilde{0}
\rangle$, in the initial and final states, $|in \rangle
\; \rightarrow |in \rangle \; |\tilde{0}\rangle$,
$|F\rangle
\; \rightarrow |F\rangle \; |\tilde{0}\rangle$,
we obtain
\begin{eqnarray}
&&
{\cal A}(|in\rangle ,|F\rangle ;T)
= - E_{in}\; E_{0} \;
({\bf d} \cdot \hat{\epsilon}_{in} ) ({\bf d} \cdot \hat{\epsilon}_{out})^{\ast}
\int_{-T/2}^{T/2} dt \int_{0}^{T/2+t}  d\tau \; 
\nonumber \\
&& \; \; \; \; \; \; \;
\sum_{{\bf k},{\bf k}'} \;
\langle F| c^{\dagger}_{{\bf k}'-{\bf k}} (t) c_{\bf k} (t-\tau) |in\rangle
\; \;
\langle \tilde{0}| \tilde{c}_{{\bf k}'} (t) \tilde{c}^{\dagger}_{{\bf k} 
+ {\bf k}_{in}} (t-\tau) 
|\tilde{0}\rangle \; \exp(i \left[ \omega_{out} t - \omega_{in} 
(t - \tau) \right]) \; .
\label{e:amp2}
\end{eqnarray}
The vacuum expectation value in (\ref{e:amp2}) is equal to
$ \langle \tilde{0}|\tilde{c}_{{\bf k}'} (t)
\tilde{c}^{\dagger}_{{\bf k}+{\bf k}_{in}} (t-\tau) |\tilde{0}\rangle$ =
$\delta_{{\bf k}',{\bf k}+{\bf k}_{in}} $ $ \times
\exp \left[ -i \left( \omega_{0} + \tilde{E}_{{\bf k}+{\bf k}_{in}}
-i \gamma/2 \right) \tau \right]$, 
where we write the energy of the excited atom as the sum of the atomic
excitation energy, $\omega_{0}$, and the kinetic energy, $\tilde{E}
_{{\bf k} + {\bf k}_{in}}$, and where $\gamma$ is the width of
the excited atomic state, $\gamma = (4/3) k^{3} d^{2}$.  
The upper limit of the $\tau$-integration interval,
may be replaced by $\infty$ if T/2+t $\gg \gamma^{-1}$.
Since we shall take the limit T $\rightarrow \infty$, we change the 
$\tau$-interval accordingly, and 
we introduce a scattering operator $\hat{S}_{\bf q}(t)$:
\begin{eqnarray}
&&
{\cal A}(|in\rangle ,|F\rangle ;T)
= - E_{in}\;  E_{0} \;
({\bf d} \cdot \hat{\epsilon}_{in} ) ({\bf d} \cdot \hat{\epsilon}_{out})^{\ast}
\int_{-T/2}^{T/2} dt \; \exp(i\omega t) \; \langle F|\hat{S}_{\bf q} (t) 
|in \rangle \; \; ,
\nonumber \\
&&
{\rm where} \; \; \; \;
\hat{S}_{\bf q} (t) = \sum_{\bf k} \int_{0}^{\infty} d\tau \;
c^{\dagger}_{{\bf k} + {\bf q}} (t) c_{\bf k}(t-\tau )
\exp(-i \left[ \omega_{0} + \tilde{E}_{{\bf k} + {\bf k}_{in}} 
- \omega_{in} -i \gamma/2 \right] \tau ) \; \; ,
\label{e:scatop}
\end{eqnarray}
where ${\bf q}$ is the momentum transfer, ${\bf q} = {\bf k}_{in} - 
{\bf k}_{out}$, and $\omega$ the energy transfer $\omega = \omega_{in} -
\omega_{out}$.
	
	The scattering rate is the ratio of the square of the amplitude
(\ref{e:scatop}) over T, summed over all $|F\rangle$-states and thermally
averaged over the $|in\rangle$-states, in the limit of T $\rightarrow \infty$.
The differential  cross-section $d^{2} \sigma_{\rm{hom}} /d\Omega
d\omega $ is the product of the resulting rate with the ratio of the
scattered particle final state density, $\rho_{out}$,
over the incident
particle flux, $J_{in}$.
For photon scattering,
$\rho_{out}=\left[ V/(2\pi)^{3}\right] \times (k^{2}/c)$, and $J_{in} =
(E_{in}^{2}/E_{0}^{2})\times (c/V)$, so that
\begin{eqnarray}
d^{2} \sigma_{\rm{hom}} /d\Omega
d\omega &=&  \lim_{T \rightarrow \infty}
\frac{1}{T} \left[ \sum_{|F>}
|{\cal A}(|in\rangle ,|F\rangle ;T)|^{2} \right]_{average (|in\rangle )}
\times \left( \rho_{out} / J_{in} \right)
\nonumber \\
&=& 
\left| (3 \gamma /4k)
(\hat{d} \cdot \hat{\epsilon}_{in}) (\hat{d} \cdot \hat{\epsilon}_{out})^{\ast}
\right|^{2} \; \frac{1}{2\pi} \; \int_{-\infty}^{\infty} dt \;
\exp(i\omega t) \; \langle \hat{S}_{\bf q}^{\dagger}(t)
\hat{S}_{\bf q}(0) \rangle \; .
\label{e:scatcros}
\end{eqnarray}
where 
$\hat{d} = {\bf d}/d$, and the $\langle \; \; \rangle$-brackets
denote the thermal 
average over the
$\langle in| |in \rangle$--matrix elements.

	To describe scattering from a dilute BEC of low depletion, 
$(N-N_{0})/N << 1 $, where
$N$ is the total number of atoms and $N_{0}$ the number of condensed atoms,
we work in the Bogoliubov approximation.
In this
scheme, we treat the zero-momentum operators as c-numbers,
replacing them by $\sqrt{N_{0}}$, 
and we keep only terms proportional to $\sqrt{N_{0}}$:
\begin{eqnarray}
\hat{S}_{\bf q}(t) \approx \sqrt{N_{0}} \times && \left[ \;
\int_{0}^{\infty} d\tau \; c^{\dagger}_{\bf q}(t) 
\exp(-i\left[ \omega_{0} + \tilde{E}_{{\bf k}_{in}} - \omega_{in} -i\gamma/2
\right] \tau )
\right.
\nonumber \\
&& \left. + \int_{0}^{\infty} d\tau \; c_{-{\bf q}}(t-\tau)
\exp(-i \left[ \omega_{0} + \tilde{E}_{{\bf k}_{in} - {\bf q}} - \omega_{in}
-i \gamma/2 \right] \tau )  \; \; \;
\right]
\label{e:sbog}
\end{eqnarray}
The first term on the right-hand side of (\ref{e:sbog}) 
represents the scattering event in which
an atom is taken out of the condensate, and the second term represents the
event in which an atom is put into the condensate, each process transferring
a momentum ${\bf q}$ to the many-boson system.  Since ${\bf k}_{in} -
{\bf q} = {\bf k}_{out}$, the excited atom momentum equals the resonant photon
wave number $k$ in both processes, and we absorb the recoil energy
$\tilde{E}_{k}$ into the
definition of the detuning: $\Delta = \omega_{in} -\omega_{0} -\tilde{E}_{k}$.

	Finally, we perform the Bogoliubov transformation to quasi-particle 
operators $b,b^{\dagger}$ :
\begin{eqnarray}
c_{\bf q}^{\dagger}(t) &=& \cosh(\sigma_{q}) \; b_{\bf q}^{\dagger}(t) \;
\; \; \; \; \; \;
\; - \; \sinh(\sigma_{q}) \; b_{-{\bf q}}(t) \; \; \; \; \; \; \; \;  ,
\nonumber \\
c_{-{\bf q}}(t-\tau) &=& \cosh(\sigma_{q}) \; b_{-{\bf q}}(t-\tau)
\; - \; \sinh(\sigma_{q}) \; b_{\bf q}^{\dagger}(t-\tau) \; \; \; ,
\label{e:bogtran}
\end{eqnarray}
where the canonical nature of the boson b-operators is ensured by
writing the coherence factors as $\cosh \sigma$ and $\sinh \sigma$.
With the time-dependence of the creation and annihilation operators,
$b_{-{\bf q}}(t-\tau) = b_{-{\bf q}}(t) \exp(i E_{q} \tau)$,
$b_{\bf q}^{\dagger}(t-\tau) = b_{\bf q}^{\dagger}(t) \exp(-i E_{q} \tau)$,
we obtain for (\ref{e:sbog}) :
\begin{eqnarray}
\hat{S}_{\bf q}(t) \approx i \; N_{0} \; \; \times &&
\left( \left[ \frac{\cosh \sigma_{q}}{\Delta + i\gamma/2} \; \;
- \frac{\sinh \sigma_{q}}{\Delta - E_{q} + i\gamma/2} \right]
\; b_{\bf q}^{\dagger}(t) \right.
\nonumber \\
&& \left. + \left[ \frac{\cosh \sigma_{q}}{\Delta + E_{q} + i\gamma/2} \; \;
- \frac{\sinh \sigma_{q}}{\Delta + i\gamma/2} \right] \; b_{-{\bf q}}(t) 
\right) \; \; , 
\label{e:sqp}
\end{eqnarray}
which describes the response of the BEC to the scattering as creating and
annihilating quasi-particles.  Each term of (\ref{e:sqp}) represents a 
different process that creates or annihilates a quasi-particle.
For example, the scattering can create a quasi-particle
($b_{\bf q}^{\dagger}$) either by 
scattering an atom from the condensate into a state of momentum ${\bf q}$
($c^{\dagger}_{\bf q}$), giving an amplitude $\cosh \sigma_{q}$ for 
quasiparticle creation ($b_{\bf q}^{\dagger}$), or by scattering an atom
from a state of momentum  $-{\bf q}$ ($c_{-{\bf q}}$) into the condensate,
giving an amplitude  $\sinh \sigma_{q}$ for quasi-particle creation.
The resonant denominators are the usual
(complex) energy-differences between the initial and intermediate state: in
the first process, the energy-difference is 
$\omega_{in} - (\omega_{0}+\tilde{E}_{k}-
i\gamma /2)$ = $\Delta + i \gamma/2$, whereas in the second process,
the intermediate state
contains an extra quasi-particle and the energy difference is equal to
$\omega_{in} - (\omega_{0}+\tilde{E}_{k}+E_{q}-
i\gamma /2)$ =  $\Delta -E_{q}+ i \gamma/2$.

	At zero temperature, 
$\langle b_{-{\bf q}}^{\dagger}(t) 
b_{-{\bf q}}(0) \rangle = 0$, 
and $\langle b_{\bf q}(t) b_{\bf q}^{\dagger}(0) \rangle = 
\exp(-iE_{q}t)$, so that
\begin{equation}
\frac{1}{2\pi}\int dt \exp(i\omega t) \langle \hat{S}^{\dagger}_{\bf q}(t)
\hat{S}_{\bf q}(0) \rangle \approx N_{0} \; \delta (\omega - E_{q}) \;
\left| \frac{\cosh \sigma_{q}}{\Delta + i \gamma/2} -
\frac{\sinh \sigma_{q}}{\Delta - E_{q} + i\gamma /2} \right|^{2}
\; \; , 
\label{e:ss}
\end{equation}
giving a spectrum that consists of a single peak.  
The scattering processes
neglected in the Bogoliubov approximation, give rise to an
additional background in the
spectrum.  The ratio of the peak intensity to the integrated background
intensity
is roughly proportional to the depletion. 
The finite-temperature generalization of (\ref{e:ss})
can be achieved in a straightforward
manner by introducing temperature-dependent quasi-particle occupation numbers
$\nu_{q}$, so that $\langle b_{-{\bf q}}^{\dagger}(t)
b_{-{\bf q}}(0) \rangle = \nu_{q} \exp(i E_{q} t)$ etc... (see for example ref.
\cite{walls}).

	The value of $\sigma_{q}$ is determined by minimizing the free
energy.  At T=0, the Bogoliubov approximation gives
$\tanh 2\sigma_{q} = \mu/(q^{2}/2m+\mu)$, where $\mu$ is the chemical potential
which, for a condensate of atoms interacting through a potential of scattering
length a, is $\mu = (4\pi /m)a\rho_{0}$ where $\rho_{0}$ is the condensate
density.
Furthermore, $E_{q} = \sqrt{(q^{2}/2m+\mu )^{2} - \mu^{2}}$, 
and with (\ref{e:scatcros}) and (\ref{e:ss}), we obtain the
following expression for the cross section
density in (\ref{e:tfinc}): 
\begin{eqnarray}
\left[
\frac{d^{2} \sigma_{\rm{inc,hom}} / d\Omega d\omega}{V} \right]
_{\mu} \; && \approx  \rho_{0} \; \delta (\omega - E_{q}) \;
\left| (3\gamma /4k)
(\hat{d} \cdot \hat{\epsilon}_{in} ) (\hat{d} \cdot \hat{\epsilon}_{out})^{\ast}
\right|^{2} \times
\nonumber \\
&&
\frac{1}{2} \left|
\frac{\sqrt{(q^{2}/2m+\mu+E_{q})/E_{q}}}{\Delta + i\gamma /2} \; \;
- \frac{\sqrt{(q^{2}/2m+\mu-E_{q})/E_{q}}}{\Delta -E_{q}+ i\gamma /2}
\right| ^{2} , 
\label{e:crosdens}
\end{eqnarray}
where $E_{q}$ and $\mu$ are ${\bf r}$-dependent.  The energy-conservation 
factor,
$\delta \left( \omega - E_{q}({\bf r})\right)$ implies
that an energy interval $(\omega , \omega + d\omega)$
probes the condensate region 
$\omega < E_{q}({\bf r}) < \omega + d\omega$.
In a magnetic trap, the excited atom experiences a potential energy due
to its magnetic moment.  We account for this effect by making the detuning
position-dependent:  $\Delta ({\bf r}) =
\Delta - \alpha V({\bf r})$, where $\alpha$ is the ratio of the excited
atom potential and the trapping potential.
Finally, for the sake of
simplicity, we specialize to a spherically symmetric harmonic oscillator
trap, $V(R) = (\omega_{T}/2) (R/L)^{2}$, where $\omega_{T}$ is the trapping
frequency and L the extent of its single-particle ground-state,
$L = 1/\sqrt{m \omega_{T}}$.  In the Thomas-Fermi approximation,
the condensate density is (\cite{GB},\cite{us})
$\rho_{0}(R) =
\left[ \mu_{T}/\frac{4\pi a}{m} \right] \times
\left[ 1 - (R/R_{0})^{2} \right] \theta (R-R_{0})$, where 
$R_{0}$ is the condensate radius, $R_{0} = L (15 a N/L)^{1/5}$ 
and $\mu_{T} = (\omega_{T}/2)(R_{0}/L)^{2}$.  In performing the spatial 
integration (\ref{e:tfinc}), spherical symmetry reduces 
the expression to an integral over the radial distance R,
and we substitute $R$ by the effective chemical 
potential $\mu$, $R = R_{0} \; \sqrt{1-\mu /\mu_{T}}$.  With this substitution,
$\rho_{0}(R) \rightarrow \mu /\frac{4\pi a}{m}$,
$\Delta(R) \rightarrow \Delta - \alpha \left[ \mu_{T} - \mu \right]$
and $\delta (\omega - E_{q}) \rightarrow \delta 
\left(\mu - \mu_{q}(\omega ) \right)
\left| \partial E_{q} / \partial \mu \right|^{-1}$, where $\mu_{q}(\omega )$
is the effective chemical potential at the positions where $E_{q}$ is
equal to $\omega$, $\mu_{q}(\omega ) = \frac{1}{2} \left[ \frac{\omega^{2}}
{q^{2}/2m} - \frac{q^{2}}{2m} \right]$, and $\frac{\partial E_{q}}
{\partial \mu} = \frac{q^{2}/2m}{E_{q}}$, we find
\begin{eqnarray}
\frac{d^{2} \sigma_{\rm{inc}}}{d\Omega d\omega} &=&
\frac{1}{\omega_{T}} \; 
\left( \frac{R_{0}^{3}}{a L^{2}} \right) \; 
\frac{\mu_{q}(\omega )}
{\mu_{T}} \; 
\sqrt{ 1-  \frac{\mu_{q}(\omega )}
{\mu_{T}}} \; 
\frac{\omega}{q^{2}/2m} \;
\left| (3 \gamma/4 k)
(\hat{d} \cdot \hat{\epsilon}_{in} ) (\hat{d} \cdot \hat{\epsilon}_{out})^{\ast}
\right|^{2} \; \times 
\nonumber \\
&& \; \; \; \; \; \; 
\frac{1}{4}
\left| \frac{\sqrt{ (q^{2}/2m+\mu_{q}(\omega ) +\omega )/\omega } }
{\Delta - \alpha \left[ \mu_{T} - \mu_{q}(\omega ) \right] + 
i\gamma /2} -  \frac{\sqrt{(q^{2}/2m+\mu_{q}(\omega ) -\omega )/\omega}}
{\Delta - \alpha \left[ \mu_{T} - \mu_{q}(\omega ) \right] 
- \omega + 
i\gamma /2} \right|^{2}
\nonumber \\
&& {\rm if} \; \; 0 < \mu_{q}(\omega) \left( = \frac{1}{2} \left[ 
\frac{\omega^{2}}{q^{2}/2m} - \frac{q^{2}}{2m} \right] \right) 
< \; \mu_{T}
\nonumber \\
&=& 0 \; \; \; \; {\rm otherwise} \; \; ,
\label{e:final}
\end{eqnarray}
The single peak at $\omega = E_{q}$
in the spectrum of the homogeneous BEC (\ref{e:crosdens}) is broadened to a
feature from $\omega = q^{2}/2m$ to $\omega = \sqrt{ (q^{2}/2m + \mu_{T})^{2}
- \mu_{T}^{2}}$.
The region in parameter-space where the difference
in intermediate state energies can be neglected,
either because
$|\omega | \ll |\Delta - \alpha ( \mu_{T} - \mu_{q}(\omega ))|$, or
$|\omega | \ll \gamma /2$ is the `fast collision' regime.
As in the off-resonant limit \cite{Jav1} \cite{Jav2},
which is part of the fast-collision
regime, fast-collision resonant scattering data
contain the same information as
non-resonant scattering data \cite{VH}, giving a cross-section proportional
the dynamical structure factor of the scattering system:
\begin{eqnarray}
&&
(\rm{in \; the \; fast \; collision \; regime \;}) \; \;
\frac{d^{2} \sigma}{d\Omega d\omega} \approx \; \;
\left| \frac{ (3 \gamma/4 k)
(\hat{d} \cdot \hat{\epsilon}_{in} ) (\hat{d} \cdot \hat{\epsilon}_{out})^{\ast}
} {\Delta - \alpha [ \mu_{T} - \mu_{q}(\omega) ] + i\gamma/2}
\right|^{2} \; S_{TF}({\bf q},\omega), \; \; \; \; \; \; \; \; \; \; \; \; \; \;
\nonumber \\
&& \; \;
\rm{where} \; \; \;
S_{TF}({\bf q},\omega) = \frac{1}{2 \omega_{T}} \left( \frac{R_{0}^{3}}{aL^{2}}
\right) \; \; \frac{\mu_{q}(\omega)}{\mu_{T}} \;
\sqrt{ 1 - \frac{\mu_{q}(\omega)}{\mu_{T}}} \; \; , \;
0 \; < \mu_{q}(\omega ) \; < \; \mu_{T} \; \; ,
\label{e:off}
\end{eqnarray}
is the dynamical structure factor of the
condensate, calculated in the above Thomas-Fermi approximation.
In Figure 1, we compare the cross-section (\ref{e:final}) and the fast-collision
approximation (\ref{e:off}) for three different values of the detuning.

        We also point out that
the dependence on the energies of the intermediate states
implies the interesting possibility of observing
the creation of a quasi-particle caused by particle annihilation.
The $\Delta - \omega + i\gamma/2$-denominator corresponds
to an intermediate state of higher energy than the
$\Delta + i\gamma/2$-intermadiate state, indicating quasi-particle
{\it creation},
whereas the intermediate state was formed by the removal
or {\it annihilation} of a boson (by exciting it to a different atomic
state). The observation of the difference in energy-denominators can
be accomplished by varying $\Delta$ while keeping $\omega$ constant, and 
requires near-resonant detuning 
which, unfortunately, can make the condensate
optically thick.
Nevertheless, experimental techniques
such as resonating on different atomic transitions and/or using the
polarization of the photons (scattered by spatially oriented dipole moments),
can reduce the optical thickness.

       Finally, to understand in what sense the scattering data can be
interpreted in a Thomas-Fermi manner, we need to investigate its
limitations.
A necessary condition for the validity of the dynamical Thomas-Fermi
description, is the validity of the static Thomas-Fermi description.
At T = 0, in a harmonic trap, a Thomas-Fermi 
condensate satisfies $R_{0} >> L$, or equivalently, 
$\mu >> \hbar \omega_{T}$
(\cite{Dalf},\cite{us},\cite{Kag}).  Furthermore, the inability of our 
dynamical 
Thomas-Fermi approach to describe the discrete spectrum of a finite-size
BEC, indicates that even for a Thomas-Fermi condensate, the dynamical
Thomas-Fermi results should be
interpreted within certain limits.
For the scattering problem, we need to realize that only fluctuations
confined to limited regions in space and time are well-described 
by a Thomas-Fermi approach.  The first constraint pertains to the excited-atom
propagation, which is only well described by means of a position dependent
detuning if, during its lifetime, the excited atom experiences a change
in potential small compared to $\hbar \gamma$,
$ (\left[ \left| {\bf F} \cdot {\bf v} \right| /\gamma \right] << \hbar
\gamma )$ where ${\bf F}$ is the force and ${\bf v}$ the velocity of the 
excited atom.  A second condition pertains to the Thomas-Fermi
description of the BEC-fluctuations.  
Consequently, we require that  $q > l_{v}^{-1}$, where  $l_{v}$
indicates the scale on which $\mu (R)$ varies
(for the harmonic oscillator trap we can choose
$l_{v} = R_{0}/3$ ; $\mu_{\rm{eff}}(R)$ varies by approximately
10 \% from $R = 0$ to $R = R_{0}/3$).  This condition is
somewhat stricter than the condition to observe incoherent
-- as opposed to coherent -- scattering, $q > R_{0}^{-1}$.
Nevertheless, with a chemical potential
$\mu_{T} \simeq 100 (\omega_{T}/2)$ (realistic
in present-day BEC technology) $R_{0} \simeq 10 L$
and $k_{c} \simeq 10/L$, where $k_{c}$ is the inverse of the local
coherence length in the middle of the trap $k_{c} = 2 \sqrt{\mu_{T} m}$.
We find then that the condition
$q > l_{v}^{-1} \simeq 3/R_{0} \simeq 0.03 k_{c}$,
leaves almost all of the interesting part of the dispersion (expected to be
phonon-like up to $k_{c}$) to be explored.  As for the
temporal constraint, the `locally homogeneous-like' excitation
picture breaks down on
the time scale that it takes the BEC-response 
to be affected by
its inhomogeneity.  Assuming that an excitation in the middle of the trap
propagates at the local speed of sound, c = $\sqrt{\mu_{T}/m}$ 
= $\omega_{T} \times R_{0} / \sqrt{2}$, we can estimate the
relevant time scale $t_{v}$ as 
$t_{v} \sim \frac{(l_{v} /R_{0})}{\omega_{T}}$.
Reducing the frequency resolution in the scattering spectrum 
to $\Delta \omega \sim t_{v}^{-1}
\sim \omega_{\rm{T}} \times (R_{0}/l_{v})$,
restricts the scattering probe to short--time
$(t < t_{v})$ `homogeneous-like' fluctuations.  Thus, the Thomas-Fermi 
scattering spectrum should be interpreted as a `smooth' version of the 
real spectrum and we should compare intensities integrated over frequency
intervals larger than or equal to $t_{v}^{-1}$ with experimental results.
These estimates were made in the middle of the trap where
the Thomas-Fermi description works best,
and which is probed on the high--frequency side 
of the cross-section, whereas the low frequency-region, $\omega \sim q^{2}/2m$,
probes the edge of the condensate where the Thomas-Fermi
results cannot be trusted.

	 P.T. was supported by Conselho Nacional de
Desenvolvimento Cientifico e Tecnologico (CNPq), Brazil.
The work of E.T. is supported by the NSF through a grant for the
Institute for Atomic and Molecular Physics at Harvard University
and Smithsonian Astrophysical Observatory.

\newpage

\underline{Figure 1}. Differential cross section for resonant
light scattering from a trap with the parameters
shown in the figure caption.  The plots show the cross section
as a function of energy transfer, at fixed momentum
transfer $q=\sqrt{m \mu_{T}}$, 
for three values of the detuning $\Delta$.
The full line shows the Thomas-Fermi calculation and the dotted
line shows the fast collision approximation result.  As the detuning
increases, the fast collision approximation becomes more and more accurate. 
 

\begin{thebibliography}{99}
\bibitem{rwa} In this formalism, we assume that the photon emission occurs 
after the photon absorption, so that we work in the rotating wave approximation.
\end{thebibliography}
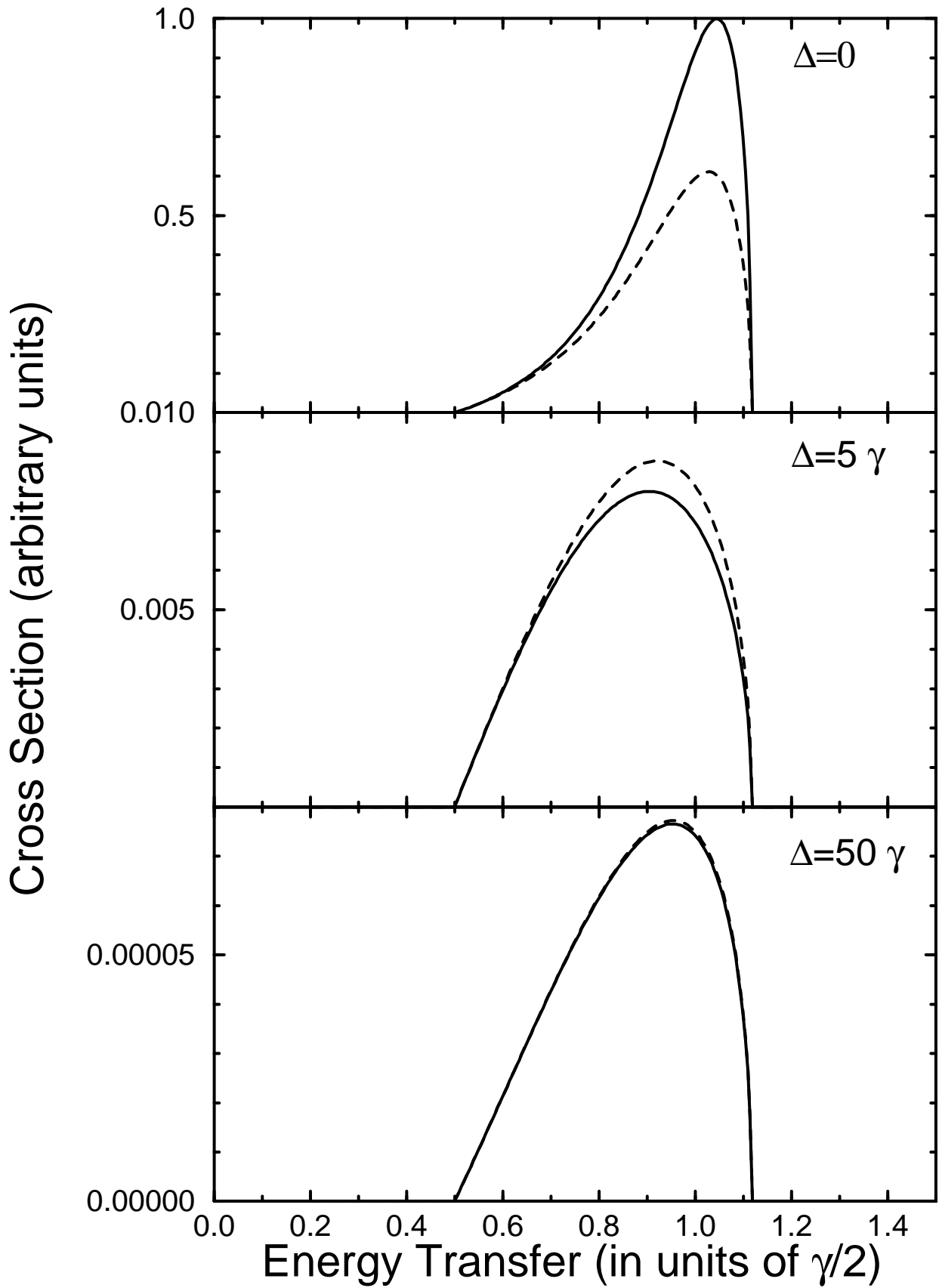
\end{document}